\documentclass[aps,twocolumn,nofootinbib]{revtex4}
\usepackage{graphicx}
\usepackage{epsfig}

\newcommand{\beq}{\begin{equation}}
\newcommand{\eeq}{\end{equation}}
\newcommand{\beqa}{\begin{eqnarray}}
\newcommand{\eeqa}{\end{eqnarray}}

\def\<{\langle}
\def\>{\rangle}

\newcommand{\complex}{{\kern .1em {\raise .47ex\hbox {$\scriptscriptstyle |$}}\kern -.4em {\rm C}}}
\newcommand{\real}{{{\rm I} \kern -.19em {\rm R}}}

\usepackage{epstopdf}

\begin{document}

\title{Quantum Communication Technology}

\author{Nicolas Gisin and Rob Thew}

\affiliation{Group of Applied Physics, University of Geneva, 1211 Geneva 4, Switzerland}

\date{\today}

\begin{abstract}
Quantum communication is built on a set of disruptive concepts and technologies. It is driven by fascinating physics and by promising applications. It requires a new mix of competencies, from telecom engineering to theoretical physics, from theoretical computer science to mechanical and electronic engineering. First applications have already found their way to niche markets and university labs are working on futuristic quantum networks, but most of the surprises are still ahead of us. Quantum communication, and more generally quantum information science and technologies, are here to stay and will have a profound impact on the XXI century.
\end{abstract}

\maketitle

{\it Introduction:}
Quantum Communication enjoys an enviable position in physics, in between fundamental quantum mechanics and applied quantum optics \cite{NPhotonics}. For most physicists, quantum communication is merely a playground to explore fascinating topics like entanglement, superposition of large objects, and, more generally, to look for places where quantum physics may fail, that is to explore the limits of quantum physics. This playground requires new technologies and concepts. Usually, new technologies are driven by applications and quantum communication is no exception: the emerging and future technologies are driven by the need for
\begin{enumerate}
\item Fast Quantum Random Number Generators (QRNG): from cryptography to internet lotteries and gaming,
\item Reliable fiber-based Quantum Key Distribution (QKD): for today's cryptography applications,
\item Quantum repeaters: for future continental scale fiber optic quantum communication,
\item Earth to satellite links: for free space quantum communication.
\end{enumerate}
The first two are already commercially available \cite{idQ} and have already found small niche markets, while the later two are still at an early stage.

{\it A Conceptual Revolution:}
The basic idea of quantum communication is to take advantage of the oddities of quantum physics, like the uncertainty relation, the superposition principle and randomness. Note the conceptual revolution: instead of being afraid of quantum peculiarities and trying to avoid their detrimental effects for standard technologies, the new generation of quantum engineers aim at exploiting the new physics. In particular they fully admit quantum physics as it stands and want to find original uses for its most counter-intuitive features. It is somewhat surprising, and disappointing, that it took six or seven decades before realizing that this new physics ought to produce new technology. One might argue that the laser, semiconductors, superconductivity, among others, are technologies based on quantum physics. However, the big difference with quantum communication - and more generally with quantum information science and technology - is that it exploits quantum physics at the level of individual quanta.

\begin{figure}[h]
\begin{center}
\epsfig{figure=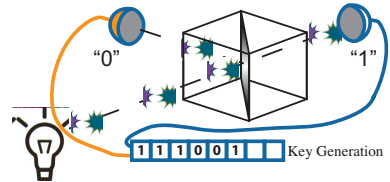,width=50mm}
\caption{Quantum random number generation - one single photon at a time is sent to a 50/50 beamsplitter and can only exit in one of the output modes. This process is fundamentally random and the photon's detection is used to the generate truly random bit strings.}
\label{fig:qrng}
\end{center}
\end{figure}

The simplest example is the quantum random number generator. Since the detection of a single photon after one of the two output ports of a beam splitter is an intrinsically random event, it offers a valuable source of randomness, see Figure \ref{fig:qrng}. Moreover, according to today's physics, such a source of randomness is unique: no device based on classical physics will ever produce true randomness, only at best "pretty good" pseudo-random numbers, or noise, whose origin is hard to fully identify. Yet, engineering a photon source, beam-splitter and two CMOS-based single-photon detectors is not that complex: the existing commercial QRNG is about the size of a match-box.

Another example, whose basic principle is rather straightforward, is QKD \cite{RMP}. Every first year quantum physics student knows that measurements tend to unavoidably disturb the quantum state of the system under investigation. This has puzzled generations of students and professors. Now, if this "negative fact" is applied to an adversary, like a spy on a communication channel where the bits are carried by quantum systems, like photons, then it is the spy who can't measure the bits without unavoidably leaving a trace of her intrusion under the form of some disturbance. Again, the reasoning is so simple that one wonders why no student came across that idea long ago (or did their professor tell them to shut-up and compute?).  Thus, while QRNGs generate randomness, QKD provides a means of distributing private (secure) randomness.

Further presentation of what quantum communication is and how it relates to entanglement and other quantum oddities can be found, e.g. in \cite{NPhotonics}. In the following we concentrate on future technology challenges.

{\it Near-Future Technologies:}
The most advanced applications of quantum communication are clearly QRNG and QKD. The first was initially developed as a component of the second. It was originally thought that QRNG would also find applications in classical cryptography and in Monte-Carlo numerical simulations. QRNG did find some application in classical cryptography (e.g. the state of Geneva uses QRNG to produce the pin-codes used for internet voting), but by far the largest application came as a surprise: internet gambling for which it has now been certified by Metas \cite{Metas} This is a good example that applications of new technologies are hard to predict (physicists are especially bad at predicting good commercial applications!).

QRNG development is a timely topic, but any approach should concentrate on three key requirements:
\begin{enumerate}
\item Origin of randomness easily identifiable. One should be able to quantify how much randomness is truly quantum and how much is "technological noise", e.g. thermal noise, detector noise, etc. 
\item Reliability, size and price. There is no fundamental reason for a QRNG not to be as small and cheap as a standard electronic chip.
\item Fast, in particular faster than classical, physical, RNG. The minimal rate of future QRNG should be in the range of hundreds of Mbps to Gbps.
\end{enumerate}

Present QKD systems are mostly based on the historical BB84 protocol \cite{BB84} (with some improvements like SARG \cite{SARGPRL04} and Decoy-state \cite{decoy1,decoy2,decoy3, decoyHugues}). However, better protocols have been invented in the context of fiber networks \cite{DPS-QKD03,COW05} . It should be understood that BB84 originally was described using polarization encoding, which is intuitively easy to understand, though in practice most real systems use some type of phase encoding that is more compatible with fiber optical systems. Despite these advances, the best QKD protocols have probably not yet been discovered. In any case, the protocol should use telecom photons (i.e. around 1550 nm), be compatible with standard optical fiber networks and combine them with the necessary quantum features to guarantee "quantum security". This requires synergy between telecom engineers and quantum theorists.

Most technology developments on QKD concentrate on single-photon sources and on detectors \cite{Hadfield10}. However, somewhat surprisingly, single-photons are not required for QKD: it is much easier to use so-called pseudo single-photon sources, i.e. strongly attenuated pulsed lasers. These are cheap, very reliable and fast (GHz rates). Note however, that single-photon sources could find their application in quantum repeaters \cite{Sangouard07}, see below. Improving single-photon detectors, on the contrary, is a real must. The best detectors in terms of efficiency are superconductor bolometers, though these are prohibitively slow and  operate at a few milliKelvin \cite{Lita08a}. QKD applications need cheap, compact, electronically cooled detectors. Today this is achieved with semiconductor detectors (InGaAs APDs) though their performance and functionality needs to greatly improve. The APDs need to have lower dark count rates and "afterpulsing" \cite{afterpulsing}, which can introduce errors on the key. Furthermore, one of the most important characteristics of single-photon detectors, that is too often neglected, is the maximum count rate; future QKD systems will need to generate several Mbps of secure keys.

Some physicists speculate that QKD systems using not-so-weak laser pulses and homodyne detection, continuous variable (CV) QKD \cite{Grangier}, will outperform single-photon schemes. They argue that, contrary to single-photon schemes, homodyne detection always produces a result. This is correct, though the results are necessarily very noisy. We expect that it is more efficient to let Nature select the cases with low noise, i.e. the cases where a single-photon is detected, rather than to always have a noisy result, where the noise has then to be removed by sophisticated error correction algorithms. Furthermore, for long distances the not-so-weak laser pulses tend to become pseudo-single-photon and the difference between the two systems vanishes. But, admittedly, the future will show us the truth with possibly both systems finding their niches.

An increasingly important requirement for future QKD schemes is that they run on the same fiber as the classical channels (both the classical processing and encrypted data channels). This is a serious challenge as the intensity difference is huge: 8 to 9 orders of magnitude. Hence, Raman scattering and other nonlinear effects have to be taken into account: even microwatts can produce enough photons to impair the quantum communication; recent efforts suggest that  multiplexing quantum and classical channels in a fiber is limited to around 50 km \cite{Walenta10} with current technology.

A serious push towards network compatibility can also be witnessed by the number of QKD test beds running or planned worldwide. In 2008 in Vienna the European consortium SECOQC demonstrated a mix of different QKD systems running in a complex network \cite{SECQOQC}. A triangular network has been running continuously in Geneva (data available in real time at www.swissquantum.com) since April 2009. In Durban, South-Africa, yet another network runs continuously carrying real data and in October 2010 a large network will commence operation in Tokyo, while others have been announced for Madrid and China.

Another sign of the maturing of QKD is the appearance of quantum hackers. They do not attack the principle on which QKD relies, as this is provable secure, but take advantage of implementation weaknesses \cite{GisinTrojan06, Makarov, Lo07}. The latter can, and have to be, tested and strengthened, rendering QKD more and more reliable.

The rate of future QKD systems will be such that true Mbps one-time pad encryption should be possible over metropolitan networks. This is including all the real-time classical processing, communication and network overheads. It will thus be the result of an interdisciplinary team of engineers.

{\it Future Technologies and Applications:}
This approach to fiber based QKD is ultimately limited in distance to a few hundred kms. Indeed, for 1000 km, even with a perfect 10 GHz single-photon source, ideal detectors and 0.2 dB/km fiber losses one would detect only 0.3 photon on average per century! Consequently, futuristic continental scale quantum communication requires completely different technologies than today's QKD systems, mere improvements will not do. There are two main paths: satellite-based  and quantum repeaters.

Satellite-based quantum communication is conceptually similar to fiber-based QKD, except that instead of fibers one sends the photons through free space between satellites and earth-based stations, both equipped with telescopes.  Since no fibers are used, the choice of wavelength is compatible with silicon APDs. The technological requirement are not so much set by quantum optics, but are defined by the stringent specifications imposed by space agencies. Currently, ESA is investigating this possibility \cite{Aspelmeyer} and the US and China also have programs in place. We wouldn't be surprised if China is the first to demonstrate a QKD link between earth and a satellite.

The alternative to satellites for continental distance quantum communication exploits a beautiful idea: quantum repeaters. Repeaters for classical optical communication, e.g. based on Erbium fiber amplifiers,  are well known, but unusable for quantum communication as any stimulated emission process necessarily brings with it spontaneous emission. At the single-photon level this noise is as strong as the signal, hence standard repeaters don't work in the quantum regime (this is a form of the quantum no-cloning theorem that follows from the linearity of Schr\"odinger's equation). The basic idea of quantum repeaters is to first establish entanglement between a series of stations, and next, to use quantum physics \cite{QtelepPRL} to teleport a "photon" (more precisely, the quantum state carried by a photon) from one station to the next. As fascinating as this is, the technological challenge is huge. First, contrary to point-to-point QKD, quantum repeaters require entanglement, i.e. the ingredient necessary for quantum teleportation. Hence, mere attenuated laser pulses will never suffice. Next, it is crucial that one distributes entanglement between neighbor stations in parallel and stores it until two neighboring stations are entangled. This required quantum memories \cite{QAPmemories}, i.e. the capacity to store photons, or more precisely their quantum state, in a reversible way without loosing any of their quantum features; in particular quantum memories ought to preserve entanglement. Furthermore, it has recently be shown that "reasonable" rates ($\ge$ 1bps over 1000 km) are possible only if the quantum memories are vastly multimode, i.e. able to store hundreds of quantum bits simultaneously.

Each of the necessary ingredient for a quantum repeater have been individually demonstrated in various labs. However, not always with compatible technology and not with sufficiently high specifications. However, there are good reasons to be optimistic. Firstly, as the challenges are truly fascinating, some of the best students are doing their PhDs on quantum repeaters. Furthermore, Europe and several national funding agencies have realized the potential and are thus supporting the research \cite{EuProjects, SwissProjects}.

{\it Conclusion:}
Quantum communication technologies involve diverse disciplines, ranging from pure engineering problems (integrate a QRNG, optimize QKD systems), to fascinating basic physics (quantum memories and teleportation), via theoretical physics (design more efficient and secure QKD protocols) and computer science (design and implement the necessary processing for the raw data delivered by QKD) and lots of electronic, telecom and software engineering (the real work). The market is still small, but burgeoning, the physics fascinating, the challenges mind-boggling. Only one thing is certain: there will be surprises.

\small

\section*{Acknowledgment}
This work has been supported by the EU projects QuReP, Q-ESSENCE and QUIE$^2$T and by the
Swiss NCCR-QP.


\begin{thebibliography}{99}

\bibitem{NPhotonics} Gisin N. and Thew, R., "Quantum Communication", Nature Photonics {\bf1}, 165-171 (2007).

\bibitem{idQ} www.idQuantique.com.

\bibitem{RMP} Gisin,~N., Ribordy,~G., Tittel,~W. and Zbinden,~H., "Quantum cryptography", {\it Rev. Mod. Phys.}, 145 (2002).

\bibitem{Metas} The Swiss national metrology institute, http://www.metas.ch.

\bibitem{BB84} Bennett,~Ch.~H. and Brassard,~ G., "Quantum cryptography: public key distribution and coin tossing", {\it Int. conf. Computers, Systems \& Signal Processing,} Bangalore, India, {\bf 10-12,} 175-179 (1984).

\bibitem{SARGPRL04} Scarani,~V., Acin,~A., Ribordy,~ G. and Gisin,~N., "Quantum cryptography protocols robust against photon number splitting attacks for weak laser pulses implementations", {\it Phys. Rev. Lett.,} {\bf 92,} 057901 (2004).

\bibitem{decoy1} Hwang,~W.-Y., "Quantum key distribution with high loss: toward global secure communication", {\it Phys. Rev. Lett.,} {\bf 91,} 057901 (2003).

\bibitem{decoy2} Wang,~X.-B., "Beating the photon-number-splitting attack in practical quantum cryptography", {\it Phys. Rev. Lett.,} {\bf 94,} 230503 (2005).

\bibitem{decoy3} Lo,~H.-K., Ma,~X. and Chen,~K., "Decoy state quantum key distribution" {\it Phys. Rev. Lett.,} {\bf 94,} 230504 (2005).

\bibitem{decoyHugues} Harrington,~ J.~W.,  Ettinger,~J.~M.,  Hugues,~R.~J. and Nordholt,~J.R., "Enhancing practical security of quantum key distribution with a few decoy states" {\it quant-ph/0503002,} Los Alamos report LA-UR-05-1156 (2005).

\bibitem{DPS-QKD03} Inoue,~K., Waks,~E. and Yamamoto,~ Y., "Differential-phase-shift quantum key distribution using coherent light", {\it Phys. Rev. A,} {\bf 68,} 022317 (2003).

\bibitem{COW05} Stucki,~D., Brunner,~N., Gisin,~N., Scarani,~V. and Zbinden,~H., "Fast and simple one-way Quantum Key Distribution", {\it App. Phys. Lett.,} {\bf 87,} 194108 (2005).


\bibitem{Hadfield10} Hadfield, R.~H., "Single-photon detectors for optical quantum information applications", Nature Phot. {\bf 3}, 696 (2009).

\bibitem{Sangouard07} Sangouard, N., Simon, C., Minar, J., Zbinden, H., de Riedmatten, H., and Gisin, N., Phys. Rev. A, {\bf 76}, 050301 (2007).

 \bibitem{Lita08a}  Lita, A.~E., Miller, A.~J., and Nam, S.~W., "Counting near-infrared single-photons with 95\% efficiency" Opt. Exp., {\bf 16}, 3032 (2008). 
 
\bibitem{afterpulsing} An afterpulse is caused by trapped charges in the APD being released when the detector is reset causing another avalanche resulting in a false detection event.

\bibitem{Grangier} Grosshan,~F. and Grangier,~Ph., "Continuous Variable Quantum Cryptography Using Coherent States",  {\it Phys. Rev. Lett.,} {\bf 88,} 057902 (2002).

\bibitem{Walenta10} Eraerds, P., Walenta, N., Legre, M., Gisin, N., Zbinden, H., "Quantum key distribution and 1 Gbit/s data encryption over a single fibre",  J. Lightwave Tech, {\bf 28}, 952 (2010)

\bibitem{SECQOQC} Peev, M., {\it et al.}, "The SECOQC quantum key distribution network in Vienna ", New J. Phys, {\bf 11} 075001 (2009).

\bibitem{GisinTrojan06} Gisin,~N., Fasel,~S., Kraus,~B., Zbinden,~H. and Ribordy,~G., "Trojan-horse attacks on quantum-key-distribution systems", {\it Phys. Rev. A,} {\bf 73,} 022320 (2006).

\bibitem{Makarov} Makarov,~V., Anisimov, A. and Skaar,~J.,  "Effects of detector efficiency mismatch on security of quantum cryptosystems", {\it Phys. Rev. A,} {\bf 74,} 022313 (2006).

\bibitem{Lo07} Qi, B.,  Fung, C.-H. F., Lo, H.-K.  and Ma, X., "Time-shift attack in practical quantum cryptosystems", Quant. Info. Compu. {\bf 7}, 43 (2007)

\bibitem{Aspelmeyer}  Aspelmeyer,~M., Jennewein,~T., Pfennigbauer,~ M., Leeb,~W. and Zeilinger,~A., "Long distance quantum communications with entangled photons using satellites", {\it IEEE J. Sel. Top. Quantum Electronics}, {\bf 9} 1541  (2005); see also the Space Quest programme at http://www.quantum.at/quest

\bibitem{QtelepPRL} Bennett,~Ch.~H., Brassard,~G., Cr\'{e}peau,~C., Jozsa,~R., Peres~A., and Wootters,~W.~K., "Teleporting an unknown quantum state via dual classical and Einstein-Podolsky-Rosen channels", {\it Phys. Rev. Lett.,} {\bf 70,} 1895 (1993).

\bibitem{QAPmemories} Simon, C., {\it et al.}, "Quantum memories: A review based on the European integrated project ÒQubit Applications (QAP)Ó",  Eur. Phys. J. D, {\bf 58} 1 (2010)

\bibitem{EuProjects} See the QuReP website at http://quantumrepeaters.eu and those listed at http://qurope.eu/projects

\bibitem{SwissProjects} For example in Switzerland, the NCCR - quantum photonics programme: http://nccr-qp.epfl.ch












\end{thebibliography}
\end{document}